\pdfoutput=1

\documentclass[usenatbib,a4paper]{mn2e}

\usepackage{amsmath}
\usepackage{amsfonts}
\usepackage{amssymb}
\usepackage{graphicx}
\usepackage[]{natbib}
\usepackage{booktabs} %change padding in table cells
\usepackage{RefJ}

\newcommand{\p}{\partial}
\newcommand{\dv}{\vec{\nabla} \cdot}
\newcommand{\x}{\vec{\nabla} \times}
\newcommand{\E}{\vec{E}}
\newcommand{\B}{\vec{B}}

\newcommand{\ddr}{d^3 \vec{r}}
\newcommand{\ddv}{d^3 \vec{v}}
\newcommand{\vr}{\vec{r}}
\newcommand{\vv}{\vec{v}}
\newcommand{\VV}{\vec{V}}

\newcommand{\VJ}{\vec{J}}

\DeclareMathOperator\erf{erf}

\title[Intergalactic Magnetogenesis at Cosmic Dawn]{Intergalactic Magnetogenesis at Cosmic Dawn by Photoionization}
\author[J-B. Durrive and M. Langer]{J-B. Durrive$^{1,2}$\thanks{E-mail:jdurrive@ias.u-psud.fr;} and M. Langer$^{1,2}$\\
$^{1}$Institut d'Astrophysique Spatiale, B\^atiment 121, Univ. Paris-Sud, UMR8617, Orsay, F-91405\\
$^2$CNRS, Orsay, F-91405}

\begin{document}

\date{Accepted ; Received ; in original form }

\pagerange{\pageref{firstpage}--\pageref{lastpage}} \pubyear{2015}

\maketitle

\label{firstpage}

\begin{abstract}
We present a detailed analysis of an astrophysical mechanism that generates cosmological magnetic fields during the Epoch of Reionization. It is based on the photoionization of the Intergalactic Medium by the first sources formed in the Universe. First the induction equation is derived, then the characteristic length and time scales of the mechanism are identified, and finally numerical applications are carried out for first stars, primordial galaxies and distant powerful quasars. In these simple examples, the strength of the generated magnetic fields varies between the order of $10^{-23}$ G on hundreds of kiloparsecs to $10^{-19}$ G on hundreds of parsecs in the neutral Intergalactic Medium between the Str{\"o}mgren spheres of the sources. Thus this mechanism contributes to the premagnetization of the whole Universe before large scale structures are in place. It operates with any ionizing source, at any time during the Epoch of Reionization. Finally, the generated fields possess a characteristic spatial configuration which may help discriminate these seeds from those produced by different mechanisms.
\end{abstract}

\begin{keywords}
  Cosmology:theory -- dark ages, reionization, first stars -- large-scale structure of Universe -- magnetic fields -- intergalactic medium.
\end{keywords}

\section{Introduction}

The origin of cosmological magnetic fields is a major open question in cosmology. Recent high energy gamma ray observations suggest that a substantial fraction, if not the whole, of the intergalactic space is magnetized \citep[e.g.][]{Neronov2010}.
The current paradigm to account for the existence of such cosmological magnetic fields states that a first mechanism, or several mechanisms combined, generated large scale magnetic fields but of very weak strengths (so-called seed fields) that were amplified later on, during structure formation, essentially through turbulence \citep[see, for instance, ][and references therein]{Brandenburg2005,Ryu2011b, Widrow2011}.
Numerous mechanisms for generating magnetic fields on cosmological scales have been proposed, operating mainly in the primordial Universe, during inflation or the electroweak and quark-hadron phase transitions \citep[e.g.][and references therein]{Grasso2001,Widrow2002, Kandus2011}. However,  mechanisms operating in the radiation dominated era \citep{Harrison1970c,Zakharov1992} or during recombination \citep[e.g.][]{Berezhiani2004,2008PhRvD..77l4028T,Fenu2011}, requiring some level of vorticity (possibly re-generated at the second order in perturbations), have also been proposed. Finally, astrophysical processes operating after recombination, capable of generating magnetic fields of cosmological interest have been investigated too. They include the well-known Biermann battery, due to a thermal electromotive force, that was first introduced in the context of stars \citep{1950ZNatA...5...65B} and later successfully considered in cosmological contexts such as structure formation \citep{Pudritz1989,Kulsrud1997,Xu2008} and the propagation of ionization fronts \citep{Subramanian1994, Gnedin2000} during Cosmological Reionization. Collision-less shocks in cosmology are also potentially capable of generating magnetic fields by triggering plasma instabilities \citep[e.g.][]{Medvedev2006,Coroniti2014}. Note that large scale magnetic fields may actually have been created  within galaxies and then transported into the intergalactic medium by powerful winds and/or jets \citep{Kronberg1999,Furlanetto2001,Beck2013}. Globally, the level at which all these mechanisms may have contributed to the magnetization of the intergalactic medium is still an open question \citep[for recent reviews, see for example][]{Widrow2011,Durrer2013,Subramanian2015} to which future observations with radio interferometers will bring essential pieces of answer \citep[e.g.][]{Bonafede2014}.

Revisiting \citet{Langer2005}, we present here a detailed analysis of an astrophysical mechanism, based on the photoionization of the Intergalactic Medium (IGM), bound to have operated during the first billion years of the Universe. \citet{Ando2010a} and \citet{Doi2011} explored numerically the same mechanism and compared it to the Biermann battery. However, in their analysis, they focused on the competition between these two mechanisms at the boundaries of self-shielded, essentially neutral clumps embedded inside the H{\sc ii} regions of individual first stars. In these conditions, they found that the Biermann battery produces stronger magnetic fields than the radiation effects, on hundreds of parsecs scales. Here, we study analytically this mechanism, relying on the momentum transfer from ionizing photons to electrons, on large scales, way outside the Str{\"o}mgren spheres of clusters of Population III stars, primordial galaxies and quasars.

In this paper, in section \ref{TheMechanism} we introduce the mechanism in full generality. Then, in section \ref{sect:oom}, we obtain a simplified order of magnitude estimate of the magnetic field strength as well as  convenient, although approximate, scaling relations. In section \ref{ApplicationCosmicDawn}, we analyse in full details the expression obtained in section \ref{TheMechanism}, apply it to the context of the Epoch of Reionization (EoR) and obtain numerical values of the magnetic fields generated in the IGM by the first luminous sources. Finally, section \ref{Discussion} is dedicated to a discussion where a comparison with the Biermann battery is also included.

\section{The Mechanism}
\label{TheMechanism}

\subsection{Presentation}

The first sources in the Universe switched on in an essentially neutral IGM, mostly made of Hydrogen, below redshifts of $30$ \citep[e.g.][]{10.2307/j.ctt24hrpv}. As they radiated, the sources formed fully ionized regions around them, called Str{\"o}mgren spheres, created mainly by photons just above the ionization threshold of 13.6 eV. However, higher energy photons were able to escape the Str{\"o}mgren spheres and propagate deeper into the IGM, because the photoionization mean free path is proportional to the cube of their energy. As pointed out by \citet{Langer2005}, these photons transferred their momentum to electrons in the surrounding, otherwise neutral medium, and thus generated radial currents. These currents were in turn able to induce large scale magnetic fields, provided the corresponding electric fields were rotational. This condition was actually satisfied thanks to the anisotropic absorption of the radiation due to the inhomogeneities of the neutral IGM.

Formally speaking, the ionization process is described microscopically as a perturbation of the distribution function of electrons. The description is then reduced to a macroscopic monofluid description to get a generalized Ohm's law, the rotational of which leads to the induction equation (section \ref{Formalism}).
The general expression for the generated magnetic field thus obtained is then examined in a simple model of the cosmological context we are interested in, using power law spectra for the sources and Gaussian profiles to model the clumpiness of the IGM. This allows us to identify the characteristic properties (characteristic length scales, typical strengths generated and field lines) of the regions that are significantly magnetized (section \ref{GlobalTerm}), and then to obtain numerical estimates of these photogenerated magnetic fields (section \ref{Results}).

\subsection[]{Formalism}
\label{Formalism}
\subsubsection{Fields}

In the non relativistic limit, displacement currents are negligible, so we consider the following Maxwell's equations:
\begin{align*}
\x \B&  =   \frac{4 \pi}{c} \VJ & \,  \dv \B& = 0\\
\x \E&  =  - \frac{1}{c}\p_t \B & \,   \dv \E& = \rho
\end{align*}
where $\VJ$ is the total current density and $\rho$ the total charge density. We will take $\rho = 0$ since the characteristic length scales of the problem are much larger than the Debye length, of the order of the kilometre here. Initially the current, the electric field and the magnetic field are null since we are interested in their ab initio generation.

Cosmological recombination was an incomplete process: during the Dark Ages, a tiny non zero ionization fraction remained in the IGM. We will call residual electrons and residual ions the free electrons and ions from this plasma. As sources switched on, they photoionized their surroundings thus liberating new electrons and ions. As we will see, it is instructive to consider separately these two types of charged species. Hence we will consider five different species, namely:
\begin{center}
$
\begin{array}{ll}
\alpha = \left\{
  \begin{array}{l}
    1: \text{residual electrons;}\\
    2: \text{residual ions;}\\
    3: \text{photoionization electrons;}\\
    4: \text{photoionization ions;}\\
    5: \text{neutrals.}
  \end{array}
    \right.
\end{array}
$
\end{center}
Each of these matter fields is characterized by its distribution function $f_\alpha$. Each $f_\alpha$ is governed by the following generalized Vlasov equation:
\begin{equation}
\p_t f_\alpha + \vv. \frac{\p f_\alpha}{\p \vr} + \frac{q_\alpha}{m_\alpha} \left(\E + \frac{\vv \times \B}{c}\right). \frac{\p f_\alpha}{\p \vv} = \left.\p_t f_\alpha \right|_c + \left.\p_t f_\alpha \right|_s,
\label{Boltzmann}
\end{equation}
where $q_\alpha$ is the charge of species $\alpha$ and $m_\alpha$ its mass. On the right hand side, the first term is the usual collision term and the second is the source term due to photoionizations that is detailed in appendix \ref{App:MomentumTerm}.

Astrophysical ionizing sources are characterized by their specific spectral density $I_\nu$. In principle one should solve the complete radiative transfer equation governing the evolution of $I_\nu$, but for our purpose it is enough to consider the following solution to this equation:
\begin{equation}
I_\nu (t, \vr, \hat{k}) = L_\nu \ \frac{e^{-\tau_\nu}}{4 \pi r^2} \delta(\hat{k}-\hat{r}),
\label{I_nu}
\end{equation}
where $L_\nu$ is the spectral luminosity density of the source and $\tau_\nu = \sigma_\nu \int_{0}^r n_{\mathrm{H\textsc{i}}} dr$ is the optical depth with $\sigma_\nu$ the photoionization cross section. The $\delta(\hat{k}-\hat{r})$ factor accounts for the fact that the source emits radially and $\frac{1}{4 \pi r^2}$ corresponds to the geometric dilution.

\subsubsection{Induction equation}

This kinetic description is convenient to build the right-hand-side of equation \eqref{Boltzmann}, but it contains much more information than needed since we are looking for the macroscopic and summed over all species quantity $\VJ$ appearing in Maxwell's equations. Thus we reduce our description to that of a monofluid \citep[cf.][and appendix \ref{App:FluidReduction}]{1973ppp..book.....K} and get the following equation governing $\VJ$, the generalized Ohm's law:
\begin{equation}
%\hspace{-0.09cm}
\begin{array}{l}
\p_t \VJ + \left(\VV . \vec{\nabla}\right) \VJ + \left(\dv \VJ\right) \VV - \VV \VV \cdot \vec{\nabla} \rho \\
%\hspace{0.5cm}
= \Sigma_\alpha \frac{q_\alpha^2 n_\alpha}{m_\alpha} \left(\E + \frac{\VV_\alpha \times \B}{c}\right) - \vec{P} + \vec{C} + \Sigma_\alpha \frac{q_\alpha}{m_\alpha} \dot{\vec{p}}_\alpha
\end{array}
\label{OhmFull}
\end{equation}
where
\begin{equation}
\dot{\vec{p}}_\alpha \equiv \int m_\alpha \vv \left.\p_t f_\alpha \right|_s \ddv
\label{sourcetermalpha}
\end{equation}

\noindent and the other terms are detailed in appendix \ref{App:FluidReduction}. The last term of equation \eqref{OhmFull} corresponds to the momentum transfer from photons to electrons.

In this paper, we are interested in the generation of magnetic fields in a cosmological context. The typical gradient scales $L$ correspond to the size of matter inhomogeneities in the high redshift intergalactic medium. To get an idea, we can consider $L \sim 10$ kpc. Further, as shown in \citet{Langer2005}, on the very short initial times typical of the plasma time-scales, the generated strengths of the fields are negligible, and the next characteristic time is set by the lifetime of the source, typically from $1$ to $100$ Myrs during Reionization. Therefore, when needed, we will take $T \sim 10$ Myrs as a typical time-scale. Finally, during Reionization the typical residual electron density is $n_e \sim 2 \times 10^{-4} \bar{n} = 4 \times 10^{-8}$ cm${}^{-3}$ at $z = 9$ \citep[e.g.][]{Seager1999}. We will use these values in the following order of magnitude estimates.

The general expression \eqref{OhmFull} may then be simplified as follows.
\begin{itemize}
\item The time variation of $\VJ$ is completely negligible here. Indeed, combining Maxwell's equations yields $\frac{E}{J} \sim \frac{L^2}{c^2 T}$ so that:
\begin{equation}
\left|\frac{\frac{q^2 n_e}{m_e} \E}{\p_t \VJ}\right| \sim \frac{q^2 n_e}{m_e c^2} L^2 \sim 10^{25} \gg 1
\end{equation}
where $q$ is the charge of an electron.
\item Since displacement currents are neglected, $\VJ$ has a rotational form from Maxwell's equations, so $\dv \VJ = 0$ here.
\item Starting from zero, $\VJ$, $\E$, and $\B$ are first order terms initially and we can therefore linearize equation \eqref{OhmFull}. Note that the generated magnetic field strengths will be small enough for this linearization to remain valid during the time-scales of interest.
\item Fluids with protons are assumed to move slowly. Thus $\VV_{\alpha}$ for ionic species are first order terms. Since $\VJ$ and $\B$ are also of first order and $m_e \ll m_i$, the second term on the right hand side of \eqref{OhmFull} is a second order term and is neglected.
\item  For the pressure term, we assume the $\dv P_\alpha^{CM}$ for protonic species ($\alpha = 2, 4,$ and $5$) is small due to their large inertia. For residual electrons, we neglect viscosity effects ensuring an isotropic pressure tensor. Photoelectrons, however, are generated radially thus introducing a small anisotropy, but we neglect this with respect to the pressure gradients of residual electrons. Hence:
\begin{equation}
\vec{P} = - \frac{q}{m_e} \vec{\nabla} p_e
\end{equation}
where $p_e$ is the residual electron pressure.
\item  For the collision term, numerous types of collisions could in principle be considered given all the species involved. However, it would prove unnecessary since all the collision frequencies are far too small and therefore the associated time-scales far too large with respect to $T$. More precisely, taking the usual linear approximation $\vec{C} = \nu_c \VJ$ where $\nu_c$ is an averaged collision frequency, and comparing this term for example to the electric field in the plasma yields
\begin{equation}
\left|\frac{\E}{\frac{m_e}{q^2 n_e} \vec{C}}\right| \sim \frac{q^2 n_e}{m_e} \frac{L^2}{c^2} \frac{\nu_c^{-1}}{T} \sim 3 \times 10^{14} \left(\frac{\nu_c}{10^{-4} \text{Hz}}\right)^{-1}.
\end{equation}
Since typical values for collision frequencies hardly exceed $10^{-4}$ Hz, this ratio is always huge.
\item  In the last term of equation \eqref{OhmFull}, the dominant and essential contribution comes from the momentum exchange between ionizing photons and photoelectrons. Other contributions are negligible because of the large inertia of protons and neutrals, and the Thomson scattering cross section which is many orders of magnitude smaller than the photoionization cross section. This last term therefore reduces to
\begin{equation}
\dot{\vec{p}}_\text{pe} = \int m_e \vv \left.\p_t f_{\text{pe}} \right|_s \ddv
\label{momentumTransfer_implicit}
\end{equation}
where `pe' stands for photoelectrons. This expression may be interpreted as follows: $\left.\p_t f_{\text{pe}} \right|_s \ddv \ \ddr \ dt$ is the number of photoelectrons generated in a volume element $\ddr$ during $dt$, appearing with momentum $m_e \vv$. Thus $\dot{\vec{p}}_\text{pe} \ \ddr \ dt$ represents the total electron momentum appearing in a volume $\ddr$ during $dt$. Equation \eqref{momentumTransfer_implicit} is a momentum density generation rate. While equation \eqref{OhmFull} has been correctly described many times as a close analogue to Newton's second law, we stress, however, that the term \eqref{momentumTransfer_implicit} is not, in essence, a force density, but a \emph{source} of momentum. Further, as detailed in appendix \ref{App:MomentumTerm}, expression \eqref{momentumTransfer_implicit} may be explicited as
\begin{equation}
\dot{\vec{p}}_\text{pe} = \frac{n_{\mathrm{H\textsc{i}}}}{c} \int_{\nu_0}^\infty f_\text{mt} (\nu) \sigma_\nu L_\nu \frac{e^{-\tau_\nu}}{4 \pi r^2} d\nu \ \hat{r}
\label{sourceterm_macro}
\end{equation}
where $f_\text{mt}$ is the fraction of momentum transferred from a photon to an electron in the photoionization process, and $\nu_0$ is the Hydrogen ionization threshold.

\end{itemize}

Finally, under these assumptions, Ohm's law \eqref{OhmFull} simplifies to:
\begin{equation}
\vec{0} = - q n_e \E - \vec{\nabla} p_e + \dot{\vec{p}}_\text{pe}.
\label{OhmSimplified}
\end{equation}
Note that $n_e$ is the total number density of electrons, but the source term $\dot{\vec{p}}_\text{pe}$ is only due to the newly photoionized electrons, and the pressure term is only due to residual electrons.
Furthermore, we emphasize again that this equation does not correspond to the balance of forces acting on single electrons \citep{Ando2010a,Doi2011}, but it rather expresses the readjustment of the electric field in the plasma in response to the apparition of new currents from photoelectrons.

The induction equation is then given by the curl of \eqref{OhmSimplified}, and may be written, using Faraday's law, as
\begin{equation}
\begin{array}{l}
\p_t \B = - \frac{c}{q} \frac{\vec{\nabla} n_e}{n_e^2} \times \vec{\nabla} p_e \\
\hspace{0.9cm} + \ \frac{c}{q n_e} \left( \frac{\vec{\nabla} x_e}{x_e} \times \dot{\vec{p}}_\text{pe} + \vec{\nabla} \int_{r_s}^r n_{\mathrm{H\textsc{i}}} dr \times \dot{\vec{q}}_{pe} \right)
\end{array}
\label{induction2}
\end{equation}
where $x_e = \frac{n_e}{n_{\mathrm{H\textsc{i}}}}$ is the total electron fraction, and $\dot{\vec{q}}_{pe}$ has the same expression as $\dot{\vec{p}}_\text{pe}$ in \eqref{sourceterm_macro} but with $\sigma_\nu^2$ instead of $\sigma_\nu$ in the integrand.
The first term on the right hand side is the usual Biermann battery term and the two additional terms are due to photoionization. The Biermann term will be discussed in section \ref{Discussion} and will not be considered here otherwise.
Then integrating \eqref{induction2}, the magnetic field at time $t$ and position $\vr$ may be written as a sum of two contributions:
\begin{equation}
\B (t, \vr) = \B_\text{local} + \B_\text{global}
\label{THE_B}
\end{equation}
where the `local' term is
\begin{align}
& \B_\text{local} = \int_0^t F_\text{local}^\text{int} \vec{F}_\text{local}^\text{geom} \ dt
\label{Blocal}\\
& F_\text{local}^\text{int} = \frac{1}{q x_e^2} \frac{1}{4 \pi r^2} \int_{\nu_0}^{\infty} f_\text{mt} \sigma_\nu L_\nu e^{-\tau_\nu} d\nu \\
& \vec{F}_\text{local}^\text{geom} = \vec{\nabla} x_e \times \hat{r}
\end{align}
and the `global' term
\begin{align}
& \B_\text{global} = \int_0^t F_\text{global}^\text{int} \vec{F}_\text{global}^\text{geom} \ dt
\label{Bglobal}\\
& F_\text{global}^\text{int} = \frac{1}{q x_e} \frac{1}{4 \pi r^2} \int_{\nu_0}^{\infty} f_\text{mt} \sigma_\nu^2 L_\nu e^{-\tau_\nu} d\nu\\
& \vec{F}_\text{global}^\text{geom} = \vec{\nabla} \left(\int_{r_s}^r n_{\mathrm{H\textsc{i}}} dr \right) \times \hat{r}.
\label{Fglobalgeom}
\end{align}
Formally, $\B_\text{local}$ and $\B_\text{global}$ are both products of two terms, integrated over time: an `interaction' term $F^\text{int}$ and a `geometric' term $\vec{F}^\text{geom}$. The interaction term characterizes the impact of the source at a time $t$ and a position $\vr$ as it includes the absorption, geometric dilution, the photoionization cross section and the fraction of momentum transferred from photons to electrons. The geometric term determines whether gradients in the IGM are indeed non radial as required, independently of the properties of the ionizing source.

Physically, one can interpret \eqref{THE_B} in the following way. The charge separation induced by photoionizations generates an electric field in the plasma satisfying the equilibrium \eqref{OhmSimplified}. For a magnetic field to grow out of it, the electric field must have a non vanishing curl.

To see how this condition may be fulfilled, consider two adjacent volume elements at a given distance from the source. In both volume elements, the equilibrium \eqref{OhmSimplified} is satisfied and dictates the value of the electric field there. Forgetting about the pressure term in this discussion, we see that $\E$ depends on two things: the local density of electrons $n_e$, and the local ability $\dot{\vec{p}}_\text{pe}$ of the source to photoionize the medium.
Thus, there are three ways for $\vec{E}$ to be different in the two volume elements: (i) they have the same $\dot{\vec{p}}_\text{pe}$ but different $n_e$, (ii) they have the same $n_e$ but different $\dot{\vec{p}}_\text{pe}$, and (iii) both $n_e$ and $\dot{\vec{p}}_\text{pe}$ are different.
Situation (i) may occur if the electron density of the plasma is locally inhomogeneous. The resulting magnetic field is the local term \eqref{Blocal}. Situation (ii) occurs when the incident distribution of photons is inhomogeneous, that is when the radiation field is anisotropic. But since the source itself emits isotropically by assumption, the only origin of such anisotropies are inhomogeneities in the medium in which photons propagate, so that absorptions along adjacent lines of sight differ. Therefore, (ii) occurs due to non radial gradients of the column densities, which gives rise to the global term \eqref{Bglobal}. Finally, situation (iii) corresponds to the general case in which everything is inhomogeneous and yields the total magnetic field \eqref{THE_B}.

Note that if at some distance two adjacent lines of sight differ, they will in general remain different further away from the source. For this reason the global term generates magnetic fields on large distances. Therefore, behind an inhomogeneity, some magnetic field is generated from this global term even if the medium is homogeneous there, and is only attenuated by geometric flux dilution, absorption and the $1/r$ factor from the gradient.

\section{Order of magnitude}\label{sect:oom}

First, to get an intuition of the efficiency of this mechanism, let us provide a crude estimate of the reached magnetic strengths  by evaluating expression \eqref{Bglobal} in the typical case of a primordial galaxy at $z = 10$. For this, as detailed in sections \ref{SourceOfIonizingPhotons} and \ref{IGM}, we consider the following parameters. We assume that the source has a Str\"omgren radius $r_s \sim 48$ kpc, a spectral index $\alpha = -2$, a lifetime $t_s = 100$ Myrs, and a total UV luminosity $L_\mathrm{tot}^\mathrm{UV} = \int_{\nu_0}^\infty L_\nu d \nu = 2.5 \times 10^7 L_\odot$. At those epochs, the ionized fraction of the IGM is $x_e \simeq 2 \times 10^{-4}$ while its mean density is $\bar{n} \simeq 2.5 \times 10^{-4}$ cm$^{-3}$. For the generation of electric fields possessing a curl, we consider that an inhomogeneity of amplitude $\delta_0 = 4$ is positioned at a distance $D \sim 1.3 r_s$ from the source. For illustration, the magnetic field strength is evaluated at a distance $r = r_s + \ell_{3 \nu_0}$ from the source, where $\ell_{3 \nu_0} = 27 \ \! \ell_{\nu_0}$ is the mean free path of photons of energy three times the Hydrogen ionization threshold.
Finally, the strength of the magnetic seed keeps on growing linearly in time as long as the momentum exchange process between photons and initially bound electrons is active. The photons we are interested in here being way above the Hydrogen ionization threshold, they do not induce any significant expansion of the Str\"omgren sphere (see discussions on $t_\mathrm{rec}$ and $t_i$ in the next section), but instead propagate outside deep into the neutral IGM. Thus, the limiting time-scale for the process to operate is essentially the life-time of the ionizing source $t_s$.

To get a simple expression of \eqref{Bglobal}, let us now take advantage of the strong dependence on $\nu$ in the optical depth to approximate absorption in the interaction term by a Heaviside function: $e^{-\tau_\nu} \simeq \theta(\nu - \nu_c)$, where $\nu_c$ is such that $\tau_{\nu_c} = 1$. Then
\begin{equation}
\nu_c (t, \vr) = \nu_0 \left(\sigma_0 \int n_\text{H\textsc{i}} dr\right)^{1/3} \simeq \nu_0 \left(\frac{r - r_s}{\ell_{\nu_0}}\right)^{1/3}
\label{nuc}
\end{equation}
where the approximation assumes a homogeneous background. Considering in addition a power law spectrum as in \eqref{powerlaw} on $[\nu_0, \infty[$ and using the fact that $r - r_s \gg \ell_{\nu_0}$ yields
\begin{equation}
F_\text{global}^\text{int} \sim \frac{8}{35} \frac{\sigma_0^2 L_\mathrm{tot}^\mathrm{UV}}{q x_e} \frac{1}{4 \pi r^2} \left(\frac{r-r_s}{\ell_{\nu_0}}\right)^{-7/3}.
\end{equation}

Finally, to estimate the contribution of the geometrical term, we consider a spherical inhomogeneity of characteristic size $\sigma$ and amplitude $\delta_0$. Then $|\vec{F}_\text{global}^\text{geom}| = \frac{1}{r} \p_\theta \int n_{\text{H\normalfont\textsc{i}}} dr$, where only the inhomogeneous part $\bar{n} \delta_0$ of the density matters, the angular variation is of the order of the angular diameter $\sigma / D$, and the integral along the line of sight is essentially the size
 of the inhomogeneity $\sigma$. Hence $|\vec{F}_\text{global}^\text{geom}| \sim \bar{n} \delta_0 \frac{D}{r}$. We may then plug in the numerical values mentioned above and get
\begin{equation}
|B|\sim 8 \times 10^{-20} \text{ G } \left(\frac{\delta_0}{4}\right) \left(\frac{L_\mathrm{tot}^\mathrm{UV}}{2.5 \times 10^7 L_\odot}\right)\left(\frac{t_s}{10^8\, \text{yrs}}\right).
\label{BOoM}
\end{equation}
This rapid estimate provides interesting information on the typical magnetic strength one might expect from this mechanism. In particular, it shows that momentum transfer from photons to electrons during Reionization is capable of producing fields with similar intensities as the usual thermal Biermann battery. However, the strength of the field strongly depends on the geometry and position of the inhomogeneity. Angularly, it is essentially governed by the geometrical term and radially by the interaction term, so that a strength of this order is reached only in specific regions.

Additionally, in accord with intuition, expression \eqref{BOoM} also shows that stronger gradients (i.e. larger inhomogeneities) will create stronger fields, and that longer-lived sources are better at creating stronger magnetic seed fields. Moreover, it indicates that the reached magnetic strength grows linearly with the UV luminosity of the ionizing source, suggesting that quasars may have produced stronger fields than faint primordial galaxies. However, the latter conclusion is actually incorrect. Indeed, as
we detail below, the dependence on the physical parameters is more involved than what equation \eqref{BOoM} suggests. For example, increasing the luminosity of the source increases not only the number of photons efficient at yielding a larger $B$, but also the number of photons at the Hydrogen ionization threshold, leading to a bigger Str\"omgren sphere (though the dependency of $r_s$ on $L_\mathrm{tot}^\mathrm{UV}$ is weak). Thus, in that case, the regions that get magnetized are actually more distant from the source, and the geometrical dilution of photons must be properly taken into account in order to get the correct estimate of $|B|$. Hence a careful and detailed analysis of \eqref{THE_B} is necessary, which is the purpose of the following section.

\section{Application At Cosmic Dawn}
\label{ApplicationCosmicDawn}

\begin{figure}
\center
\includegraphics[scale=0.7]{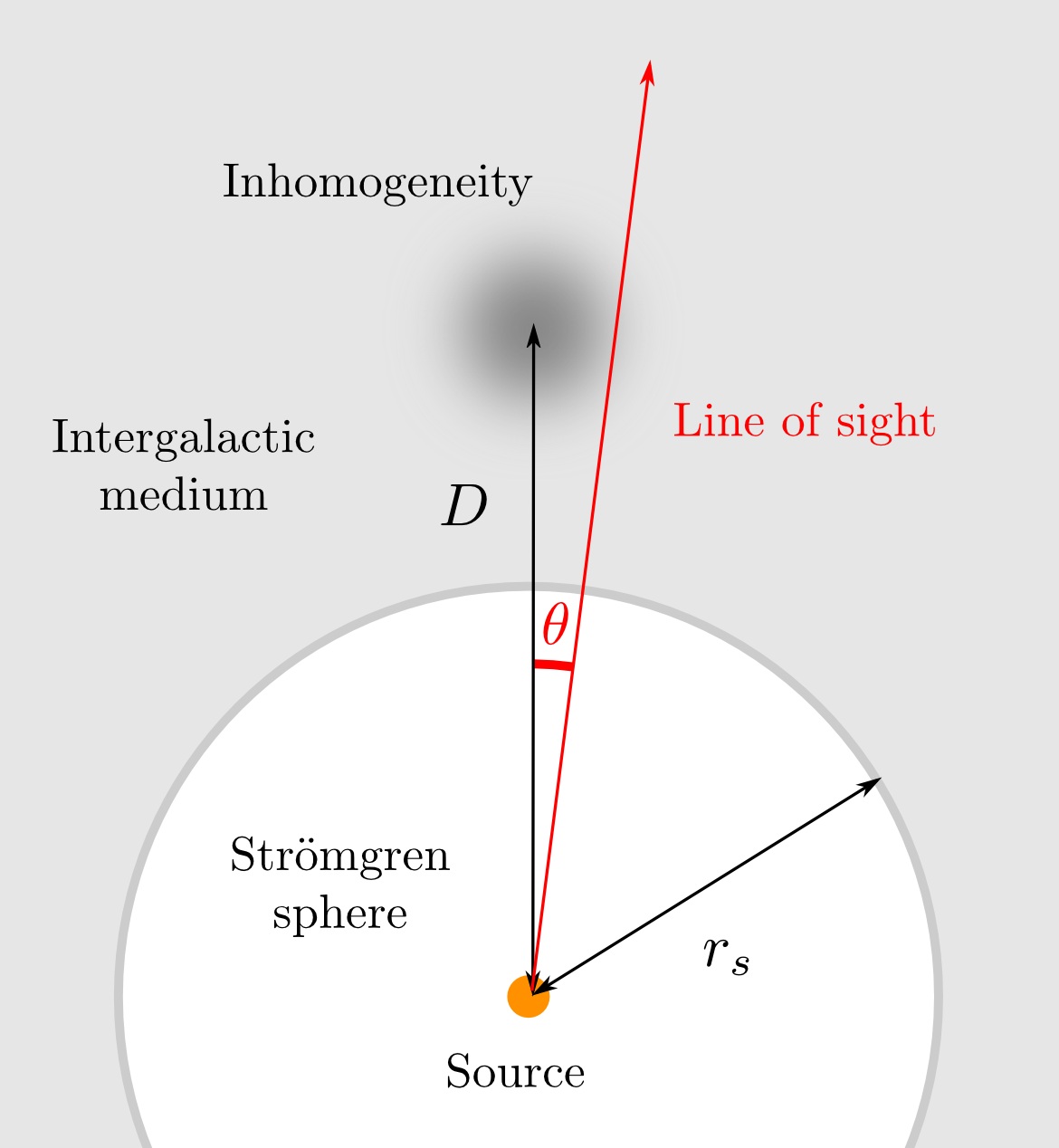}
\caption{Representation of the modelled situation. A source forms a Str{\"o}mgren sphere of radius $r_s$. An inhomogeneity in the IGM is situated at a distance $D$ from the source. All the graphs in this paper are plotted along the line of sight corresponding to $\theta = \theta_{\text{max}}$ defined by equation \eqref{thetamax}.}
\label{Schema_Situation}
\end{figure}

The analytical formula of the magnetic field \eqref{THE_B} we obtained is in principle applicable during Reionization in the vicinity of any ionizing source embedded in an arbitrarily inhomogeneous neutral medium. To gain further insight and obtain numerical estimates, we now apply this expression to some simple models, considering specific sources and a mildly non-linear inhomogeneity outside their Str{\"o}mgren sphere, as represented in figure \ref{Schema_Situation}.

\subsection{Sources of ionizing photons}
\label{SourceOfIonizingPhotons}
We consider the three different types of sources that are believed to have driven Reionization: (clusters of) Population III stars, primordial galaxies, and high redshift quasars (QSOs).
For simplicity, each source is characterized by a power law luminosity, with normalisation $L_0$ and spectral index $\alpha$ in a certain frequency range:
\begin{equation}
L_\nu = L_0 \left(\frac{\nu}{\nu_0}\right)^\alpha \text{ for } \nu \in [\nu_0,\nu_1]
\label{powerlaw}
\end{equation}
where we will call $\nu_1$ the cut-off frequency. For quasars, we take $\alpha \sim -1.7$, $\nu_1 = 100 \nu_0$ and a total luminosity of $10^{12} L_\odot$ which fixes $L_0$ \citep[e.g.][]{Shang2011}.
Values for the other sources were based on the currently available synthetic spectra from the Yggdrasil model\footnote{http://ttt.astro.su.se/$\sim$ez/} \citep{Zackrisson2011}, which uses the \cite{Schaerer2001} and \cite{Raiter2010} single stellar populations for population III stars, differing by their initial mass functions (IMF). We used the PopIII.1 model for Population III clusters, a zero-metallicity population with an extremely top-heavy IMF ($50-500 M_\odot$, Salpeter slope), from which we obtained $\alpha \sim - 0.3$, $\nu_1 = 4 \nu_0$ and $L_0 = 10^{20}$ erg s$^{-1}$ Hz$^{-1}$. For primordial galaxies we considered Population II stars with a Kroupa IMF and metallicity $Z \in [0.0004,0.008]$, which yields $\alpha \sim - 2$, $\nu_1 = 4 \nu_0$ and $L_0 = 3 \times 10^{25}$ erg s$^{-1}$ Hz$^{-1}$. In practice, we checked that our results do not depend on metallicity.
From \citet{Martini2004}, we assumed quasars to live about $100$ Myrs. For Pop III clusters and first galaxies, we considered constant star formation rates for $100$ Myrs.

For simplicity the Str{\"o}mgren spheres of the sources are taken spherically symmetric. In reality, they are far from symmetric, exhibiting often a `butterfly' shape in numerical simulations \citep[e.g.][]{2001MNRAS.324..381C}, so that $r_s = r_s(\theta,\varphi)$ in principle. In this case, the lower boundary of the integral defining the column density in equation \eqref{Fglobalgeom} varies spatially. In this first approach we neglect these angular variations.
The magnitude of the radius of Str{\"o}mgren spheres depends on whether the recombination rate is sufficient to reach the steady state. The recombination time-scale is $t_{\text{rec}} = (\alpha_B \bar{n} C)^{-1}$, where $\bar{n}$ is the mean hydrogen density, $\alpha_B = 2.6 \times 10^{-13}$ cm$^3$ s$^{-1}$ is the case B recombination coefficient at a gas temperature of $10^4$ K and $C(z)$ is the hydrogen clumping factor. This factor is still poorly constrained, and  we adopt the fit  $C(z) = 27.466 \exp(-0.114 z + 0.001328 z^2)$ obtained from simulations by \citet[][]{2006MNRAS.372..679M}. We note that the  values of the clumping factor are somewhat sensitive to its definition \citep[e.g.][]{2012MNRAS.427.2464F}, but they remain of the same order of magnitude. In our redshift range, this yields $t_{\text{rec}} \sim 10$ Myrs while our sources live $100$ Myrs so, for simplicity, we considered the following expression of the Str{\"o}mgren radius
\begin{equation}
r_s = \left(\frac{3 \dot{N}_{\text{ion}}}{4 \pi \alpha_B C \bar{n}_{\mathrm{HI}}^2}\right)^{1/3},
\label{StromgrenRadius}
\end{equation}
where the rate of ionizing photons emitted by the source is $\dot{N}_{\text{ion}} = \int_{\nu_0}^\infty \frac{L_\nu}{h \nu} d\nu$.
Finally, outside Str{\"o}mgren spheres, Thomson scattering is negligible since in neutral regions its cross section is by numerous orders of magnitude smaller than that of photoionization.

\subsection{Intergalactic medium}
\label{IGM}
For simplicity, we neglect the contribution of the first Helium ionization and we consider a homogeneous Reionization scenario. We thus suppose that the ionization contrast $\delta_x$ vanishes, meaning that the ionization fraction $x_e$ is uniform, so that $\vec{\nabla} x_e = \vec{0}$. In the rest of the paper, we will therefore focus only on the global term in equation \eqref{THE_B}.\\
The dynamical behaviour of baryons in the IGM is governed by
\begin{equation}
\frac{dn_e}{dt} = \Gamma_p n_{\mathrm{H\textsc{i}}} - \alpha_B n_e^2 = - \frac{dn_{\mathrm{H\textsc{i}}}}{dt}
\label{dnHdt}
\end{equation}
where the photoionization rate at a distance $\vr$ from the source is
\begin{equation}
\Gamma_p(t, \vr) = \frac{1}{4 \pi r^2} \int_{\nu_0}^\infty \frac{\sigma_\nu}{h \nu} L_\nu e^{-\tau_\nu} d\nu.
\label{Gamma_p}
\end{equation}
Outside the Str{\"o}mgren sphere, since recombinations are negligible, the variation time-scale of the densities $n_{\mathrm{H\textsc{i}}}$ and $n_{\mathrm{e}}$ is the ionization time-scale $t_i = \Gamma_p^{-1}$, which depends on the distance from the source. These densities may be considered constant wherever $t_i/t_s$ is smaller than one. The distance at which $t_i = t_s$ is indicated by a vertical dashed line in figures \ref{OverVsUnderdensity}, \ref{Figures_Spectra} and \ref{LapinCretin}. To the right of this line, the assumption of constant $n_{\mathrm{H\textsc{i}}}$ and $n_{\mathrm{e}}$ is perfectly safe. Between $r_s$ and this distance, their variation is not negligible in principle, but note that this line lies, in the vast majority of cases, very close to the Str{\"o}mgren radius. Therefore the assumption of constant $n_{\mathrm{H\textsc{i}}}$ and $n_e$ does not affect significantly the values of the magnetic field beyond this line.

To model simply the inhomogeneous IGM in which the source is embedded, we consider an inhomogeneity next to the Str{\"o}mgren sphere, centred at a position $\vec{D}$ from the source (cf. figure \ref{Schema_Situation}), with various profiles. The first simple profile consists in a Gaussian inhomogeneity
\begin{equation}
n_{\mathrm{H\textsc{i}}} = \bar{n} \left(1 + \delta_0 e^{- \frac{(\vr-\vec{D})^2}{2 \sigma^2}}\right),
\label{GaussianInhomogeneity}
\end{equation}
which is an overdensity for $\delta_0 > 0$ and an underdensity for $-1 \leq \delta_0 < 0$. Such a Gaussian overdensity may result from gravitational instability, but inhomogeneities may also form through thermal instability, collecting the surrounding matter, in which case the overdensity is surrounded by an underdense region. Hence a second simple model of inhomogeneity, which we will call hereafter a `Mexican hat' profile (MH), composed of two imbricated Gaussians, a small width overdensity $\delta_0^+ > 0$ inside an extended width underdensity $\delta_0^- < 0$:
\begin{equation}
n_{\mathrm{H\textsc{i}}} = \bar{n} \left(1 + \delta_0^+ e^{- \frac{(\vr-\vec{D})^2}{2 \sigma_+^2}} + \delta_0^- e^{- \frac{(\vr-\vec{D})^2}{2 \sigma_-^2}}\right).
\label{MHInhomogeneity}
\end{equation}
Mass conservation $\int \left(n_{\mathrm{H\textsc{i}}} - \bar{n}\right) d^3r = 0$ then gives the constraint $\delta_0^+ \sigma_+ + \delta_0^- \sigma_- = 0$.

\subsection{The global term}
\label{GlobalTerm}

\bgroup %change padding in table cells
\def\arraystretch{1.5} %change padding in table cells

\begin{table}
\begin{minipage}{86mm}
\begin{center}
$\begin{array}{c|c|c|c}
z & \ell_{\nu_0} \text{(kpc)} & \ell_{4 \nu_0} \text{(kpc)} & \ell_{10 \nu_0} \text{(kpc)} \\
\hline
30 & 0.0073 & 0.47 & 7.3 \\
15 & 0.053 & 3.4 & 53 \\
10 & 0.16 & 11 & 160 \\
6 & 0.64 & 41 & 640 \\
\end{array}$
\end{center}
\caption{\label{TableMfp} Orders of magnitude of mean free paths of photons of frequency $\nu = \nu_0, 4 \nu_0$ and $10\nu_0$ at
various redshifts during the EoR.}
\end{minipage}
\end{table}

In order to analyse in detail expression \eqref{Bglobal} and explicit its characteristics, we now consider an inhomogeneity with a Gaussian profile together with a source with a power law spectrum.
Given the importance of the mean free path of photons in this model, we recall some typical orders of magnitude in table \ref{TableMfp}, for frequencies typical of the ionizing photons emitted by the sources participating to the EoR.

\begin{figure*}
\includegraphics[scale=0.6]{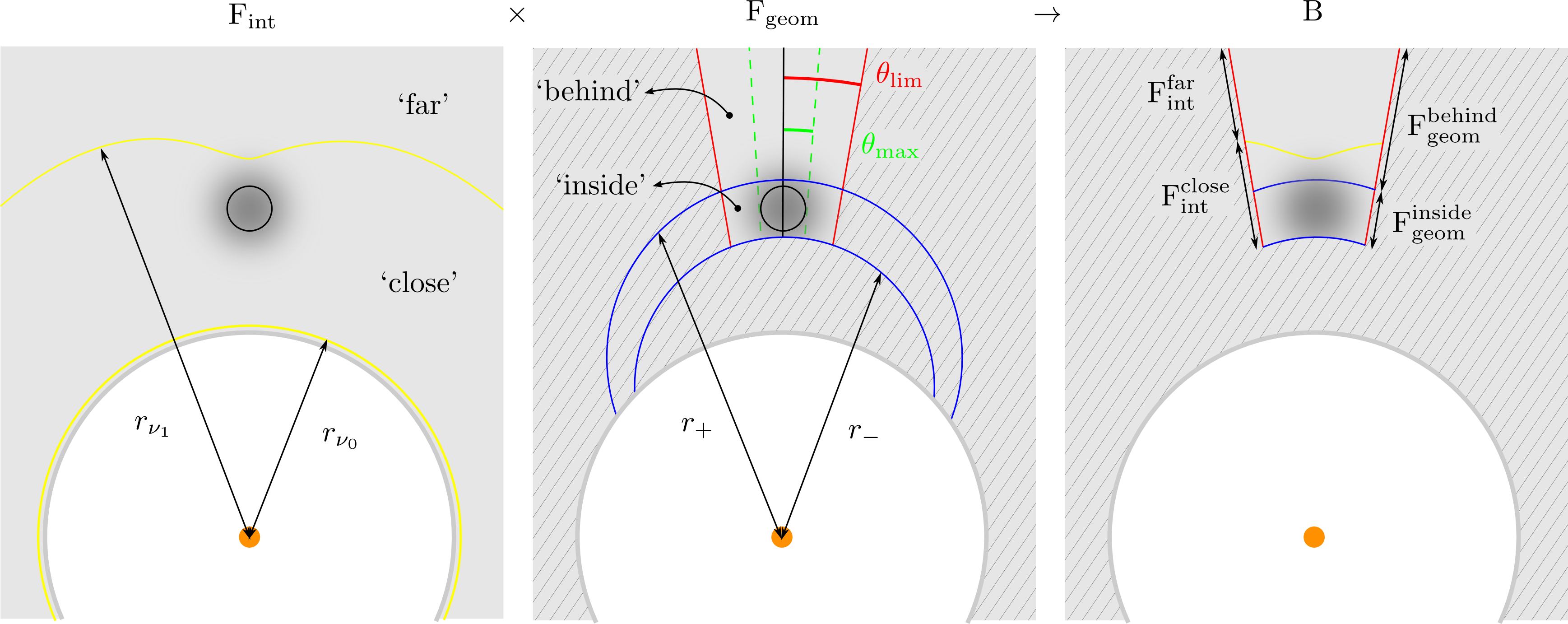}
\caption{The properties of the magnetic field depend on whether the inhomogeneity is `close' or `far' from the Str{\"o}mgren sphere ($r \ll r_{\nu_1}$ or $r_{\nu_1} \ll r$, left panel), and on whether we are considering the regions `inside' or `behind' the inhomogeneity ($r_{-} \ll r \ll r_{+}$ or $r_{+} \ll r$, central panel). Thus, four different zones may be distinguished in which the magnetic field is the most significant (right panel). Hatching indicates regions where the strengths are weaker.}
\label{Regimes}
\end{figure*}

\subsubsection{The interaction term}
As we have seen in equation \eqref{Bglobal}, the strength of the magnetic field depends on two criteria: the intensity of the interaction of photons at a given distance, quantified by $F_\text{int}$, and on the geometry of the situation, quantified by $F_\text{geom}$. Let us first focus on the former.
For a power law spectrum \eqref{powerlaw}, with the change of variable $y = \tau_\nu$, the interaction term reads
\begin{equation}\label{InteractionTerm}
  \begin{array}{l}
    \text{F}_\text{int} = \xi(r) \left( \frac{\gamma(\frac{6 - \alpha}{3}, \tau_{\nu_1}) - \gamma(\frac{6 - \alpha}{3}, \tau_{\nu_0})}{\tau_{\nu_0}^{\frac{6 - \alpha}{3}}} \right.\\
    \hspace{1.5cm} - \left. \frac{\gamma(\frac{5 - \alpha}{3}, \tau_{\nu_1}) - \gamma(\frac{5 - \alpha}{3}, \tau_{\nu_0})}{\tau_{\nu_0}^{\frac{5 - \alpha}{3}}} \right)
  \end{array}
\end{equation}
where $\gamma$ is the lower incomplete gamma function. To lighten the expressions we have set $\xi (r) \equiv \frac{1}{q x_e} \frac{8}{5} \frac{1}{4 \pi r^2} \frac{(\sigma_0)^2 L_0 \nu_0}{3}$ and dropped the `global' exponent from now on.

Now define $r_{\nu_0}$ the distance from the source at which $\tau_{\nu_0} = 1$, and $r_{\nu_1}$ such that $\tau_{\nu_1} = 1$. Note that they roughly correspond to the mean free paths $\ell_{\nu_i}$ from the Str{\"o}mgren sphere, $r_{\nu_i} \sim r_s + \ell_{\nu_i}$, with equality for $\delta_0 \ll 1$.
Then, three regions may be defined in the interaction term (cf left panel of figure \ref{Regimes}) namely: `very close' to ($r_s \leq r \ll r_{\nu_0} \ll r_{\nu_1}$), `close' to ($r_{\nu_0} \ll r \ll r_{\nu_1}$) and `far' from ($r_{\nu_0} \ll r_{\nu_1} \ll r$) the Str{\"o}mgren sphere. Given the smallness of $\ell_{\nu_0}$ (cf table \ref{TableMfp}), the `very close' regime is not relevant for our purpose here as it lies within the width of the Str{\"o}mgren radius.

In the `close' region, the interaction term \eqref{InteractionTerm} simplifies to
\begin{equation}
\text{F}_\text{int}^\text{close} = \xi(r) \left( \frac{\Gamma\left(\frac{5 - \alpha}{3}\right)}{\tau_{\nu_0}^{\frac{5 - \alpha}{3}}} - \frac{\Gamma\left(\frac{6 - \alpha}{3}\right)}{\tau_{\nu_0}^{\frac{6 - \alpha}{3}}} \right)
\label{Fintclose}
\end{equation}
while `far' from the Str{\"o}mgren sphere
\begin{equation}
\text{F}_\text{int}^\text{far} = \xi(r) \left(1 - \frac{\nu_0}{\nu_1}\right) \left( \frac{\nu_1}{\nu_0}\right)^{\alpha - 5} \frac{e^{- \tau_{\nu_1}}}{\tau_{\nu_1}}.
\label{Fintfar}
\end{equation}
In other words, the source has a certain `impact zone' (the ̀`close' region) inside which its ionizing photons are numerous enough to interact significantly. Outside this region, the strength of the field decreases exponentially.

\subsubsection{The geometric term}
Let us now focus on the second factor of expression \eqref{Bglobal}. To see precisely what the requirement of a favourable geometrical condition consists of, let us consider the simple case of a Gaussian inhomogeneity \eqref{GaussianInhomogeneity}. In spherical coordinates as in figure \ref{Schema_Situation}, the geometric term reads
\begin{align}
 \text{F}_\text{geom}  = & - \frac{\sqrt{2} \sigma \bar{n} \delta_0}{r} \ F_1 \left(\sin \theta \frac{D}{\sigma}\right) \nonumber\\
& \times \left[F_2 \left(\frac{r - \cos \theta D}{\sigma}\right)-F_2 \left(\frac{r_s - \cos \theta D}{\sigma}\right)\right]
\end{align}
where
\begin{equation}
F_1(x) \equiv \frac{x}{\sqrt{2}} \ e^{-\frac{x^2}{2}}
\end{equation}
and
\begin{equation}
F_2(x) \equiv e^{-\frac{x^2}{2}} - \sqrt{\frac{\pi}{2}} \cos \theta \frac{D}{\sigma} \erf \left(\frac{x}{\sqrt{2}}\right).
\end{equation}

As far as the angular dependence is concerned, it is dominated by the $F_1$ factor which is itself characterized by the two angles:
\begin{align}
& \theta_{\text{max}} = \arcsin \left(\frac{\sigma}{D}\right)
\label{thetamax}\\
& \theta_{\text{lim}} = \arcsin \left(\frac{3 \sqrt{3}}{2} \frac{\sigma}{D}\right)
\end{align}
where $\theta_{\text{max}}$ is the angle corresponding to the maximum of the function $F_1(\sin \theta \frac{D}{\sigma})$ and $\theta_{\text{lim}}$ is the angle for which the tangent at the inflexion point goes to zero. For angles larger than $\theta_{\text{lim}}$, the strength of the interaction term is very strongly decreasing so that we will only consider lines-of-sight such that $\theta \in [-\theta_{\text{lim}},\theta_{\text{lim}}]$.

For the radial dependence, defining
\begin{equation}
r_{\pm}(\theta) = \cos \theta D \pm \sqrt{2} \sigma,
\end{equation}
three regions may be distinguished (cf. central panel of figure \ref{Regimes}) namely: `in front of' ($r_s \leq r \ll r_{-}$), `inside' ($r_{-} \ll r \ll r_{+}$) and `behind' ($r_{+} \ll r$) the inhomogeneity. `In front of' the inhomogeneity, the two $F_2$ terms essentially cancel each other out, so that $\B \sim \vec{0}$ in this region.
`Inside' the geometric term simplifies to
\begin{equation}
\text{F}_\text{geom}^\text{inside} = \frac{\bar{n} \delta_0 D^2}{2 \sigma^2} e^{- \frac{\left( \sin \theta D\right)^2}{2 \sigma^2}} \sin 2 \theta \left(1 - \cos \theta \frac{D}{r} + \sqrt{\frac{\pi}{2}} \frac{\sigma}{r}\right),
\label{Fgeom1}
\end{equation}
while `behind' the inhomogeneity
\begin{equation}
\text{F}_\text{geom}^\text{behind} = \bar{n} \delta_0 \sqrt{2 \pi} \sin \theta \frac{D^2}{r \sigma} e^{- \frac{\left( \sin \theta D\right)^2}{2 \sigma^2}}.
\label{Fgeom2}
\end{equation}

In other words, this geometrical factor decreases the strength of the field exponentially above a certain angle $\theta_\text{lim}$. It hence constraints the magnetic field to be generated only in the vicinity and behind inhomogeneities, within a domain in the shape of a shadow.

\subsubsection{Magnetized regions}

The result is that the field has a rather simple spatial distribution: It reaches its maximum strength in the vicinity of the inhomogeneity, roughly at $\theta = \theta_{\text{max}}$ and $r = D$, and then decays with distance behind the inhomogeneity (cf figure \ref{Gaussian}).
More precisely, to a given source corresponds a given $r_{\nu_1}$ delimiting its `close' and `far' regions. Then the properties of the magnetic field generated around this source essentially depend on whether the surrounding inhomogeneities are inside the `close' or `far' regions as depicted on the right panel of figure \ref{Regimes}.

Then we can see that if the inhomogeneity is close to the source ($D < r_{\nu_1}$), the strength reaches
\begin{align}
B_\text{close}^\text{inside}  =\ & t \ \xi \! \left(D\right) \bar{n} \delta'_0 \left[\Gamma\left(\frac{5-\alpha}{3}\right) \left(\frac{D - r_s + \delta'_0 \sigma}{\ell_{\nu_0}}\right)^{(\alpha-5)/3}\right. \nonumber \\
& \quad \left. - \Gamma\left(\frac{6-\alpha}{3}\right) \left(\frac{D - r_s + \delta'_0 \sigma}{\ell_{\nu_0}}\right)^{(\alpha-6)/3}\right],
\label{Bcloseinside}
\end{align}
while if it is far from it ($D > r_{\nu_1}$), it reaches
\begin{equation}
B_\text{far}^\text{inside} = t \ \xi \! \left(D\right) \bar{n} \delta'_0 \left(1 - \frac{\nu_0}{\nu_1}\right) \left(\frac{\nu_0}{\nu_1}\right)^{5-\alpha} \frac{\exp\left(- \frac{D - r_s + \delta'_0 \sigma}{\ell_{\nu_1}}\right)}{\frac{D - r_s + \delta'_0 \sigma}{\ell_{\nu_1}}}
\label{Bfarinside}
\end{equation}
where $\delta'_0 \equiv \sqrt{\frac{\pi}{2 e}} \delta_0$. Then after reaching these values, the field decays behind the inhomogeneity depending on the situation according to
\begin{equation}
B_\text{close}^\text{behind} \propto r^{-3} \left(\frac{r - r_s + 2 \delta'_0 \sigma}{\ell_{\nu_0}}\right)^{\frac{\alpha - 5}{3}}
\label{Bclosebehind}
\end{equation}
or
\begin{equation}
B_\text{far}^\text{behind} \propto r^{-3} \frac{\exp\left(- \frac{r - r_s + 2 \delta'_0 \sigma}{\ell_{\nu_1}}\right)}{\frac{r - r_s + 2 \delta'_0 \sigma}{\ell_{\nu_1}}}.
\label{Bfarbehind}
\end{equation}
In these expressions we can see explicitly the role played by the amplitude of the inhomogeneity $\delta_0$. First of all, the sign of $B$ is given by the sign of $\delta_0$. Also, we can see that the strengths are larger in underdensities than in overdensities. Indeed, for negative $\delta_0$, $D - r_s + \delta'_0 \sigma$ is smaller than for a positive $\delta_0$. Expressions \eqref{Bcloseinside} and \eqref{Bfarinside} thus yield larger values for a negative $\delta_0$ than a positive one. The same applies to expressions \eqref{Bclosebehind} and \eqref{Bfarbehind}, showing that the fields will also be generated over larger distances in underdensities than in overdensities.

\subsection{Sensitivity to physical parameters}
\label{Results}

Figure \ref{Gaussian} shows a typical result in the case of an underdensity, for a primordial galaxy at $z=10$. The superimposed lines are separating the different regions studied in the above section. Interesting field strengths are produced in regions that span transversally the domain between $- \theta_\text{lim}$ and $\theta_\text{lim}$, and extend radially over a few $\ell_{4 \nu_0}$ behind the inhomogeneity. In addition, the dashed red arc represents the distance corresponding to $t_i = t_s$, below which $n_{\mathrm{H\textsc{i}}}$ is overestimated as discussed in section \ref{IGM}.
\begin{figure}
\includegraphics[scale=0.3]{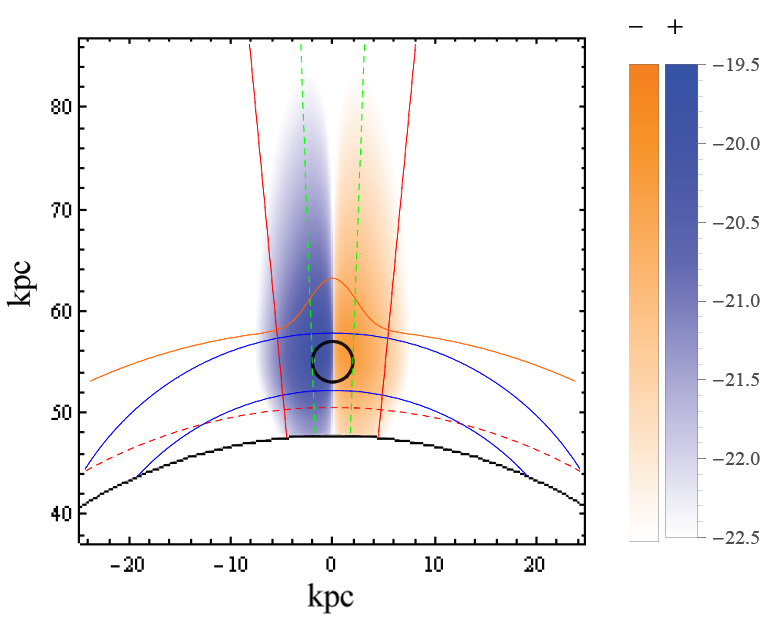}
\caption{Typical field generated by an underdensity at $z=10$ by a primordial galaxy. The thin black arc represents the edge of the Str{\"o}mgren sphere. The thick black circle corresponds to the FWHM of the underdensity. The blue area corresponds to magnetic fields pointing towards the reader, orange in the opposite direction. Strength is indicated in Gauss in logarithmic scale. Corresponding to the analytical decomposition explicited in figure \ref{Regimes}, the continuous red lines show the $- \theta_\text{lim}$ and $\theta_\text{lim}$ directions, the dashed green lines correspond to the $- \theta_\text{max}$ and $\theta_\text{max}$ directions, the blue arcs indicate $r_{-}$ and $r_{+}$, and the continuous orange curve indicates $r_{\nu_1}$ ($\nu_1 = 4 \nu_0$ here). The dashed red arc is the distance at which $t_i = t_s$ (cf text).}
\label{Gaussian}
\end{figure}
The strength and scales reached by the generated field depend on many parameters: those characterizing the IGM ($D$, $\sigma$ or $\sigma_+$, $\delta_0$ or $\delta_0^+$ and $\bar{n}$), those characterizing the source ($L_0$, $\alpha$ and $\nu_1$), and the redshift $z$. In light of the above analytical expressions, we now make those parameters vary one by one in order to evaluate their importance.

\subsubsection{Varying the properties of the inhomogeneity}
For simplicity all the configurations (source and inhomogeneity) considered here possess an axial symmetry (cf figure \ref{Schema_Situation}). Therefore there are no orthoradial gradients along the symmetry axis, and thus the magnetic field vanishes in the $\theta = 0$ direction. For this reason, all the graphs presented here are plotted along the line of sight $\theta = \theta_{\text{max}}$. Note also that all graphs start at $r = r_s$.
Obviously, the larger $|\delta_0|$ the larger the gradients, and thus the higher the strengths generated in principle. On the other hand, large $\delta_0$ imply strong absorption in the case of an overdensity. So for example behind a very dense region, the field vanishes but is important at its edges. In the following, we consider pockets of neutral gas, not virialized, in the vicinity of an early isolated ionizing source. Therefore we consider $- 1 \leq \delta_0 \lesssim 5$.

In figure \ref{OverVsUnderdensity} we plot the field generated in an overdensity and an underdensity, differing only by the sign of $\delta_0$. We see that the strength is higher and generated on larger scales in the underdensity case, which is due to the fact that there are more photons at a given distance than in the overdensity case. This was already noticed previously from expressions \eqref{Bcloseinside} to \eqref{Bfarbehind}.

\begin{figure}
\includegraphics[width=84mm]{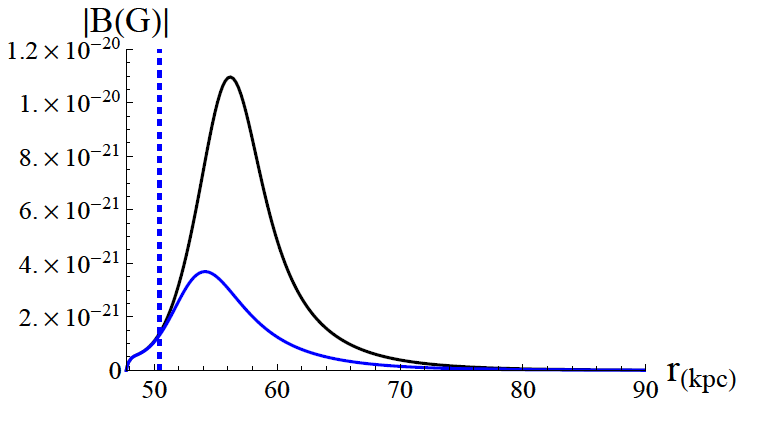}
\caption{Magnetic field generated by a primordial galaxy, with two Gaussian inhomogeneities of same $|\delta_0|$, but with opposite signs (blue: positive; black: negative). We see that the mechanism is more efficient in underdense regions. Vertical dashed line indicates the distance at which $t_i = t_s$.}
\label{OverVsUnderdensity}
\end{figure}
In figure \ref{Figures_Inhomo} we study the influence of the position and profile of the inhomogeneity for a given source. For illustrative purposes we take a QSO at redshift $z = 10$.
\begin{figure*}
\includegraphics[scale=0.27]{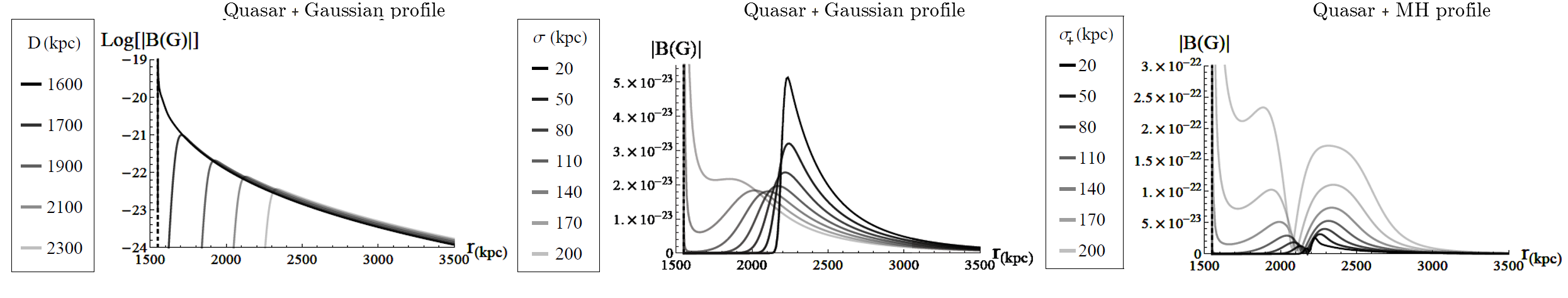}
\caption{Magnetic field generated with various inhomogeneities, for a given source (here a quasar at $z=10$). Left: Gaussian overdensity of fixed profile, varying distance. Middle: Gaussian overdensity at fixed position, varying its width $\sigma$. Right: Mexican hat profile at fixed position, varying its width $\sigma_+$. The lines corresponding to distances at which $t_i = t_s$ (cf. section \ref{IGM}) are not visible here because for quasars the Str{\"o}mgren sphere is so large that photons are very diluted as they reach $r_s$, and $t_i$ becomes much larger than $t_s$ already at the edge of the Str{\"o}mgren sphere.}
\label{Figures_Inhomo}
\end{figure*}
On the left of the figure, the profile of the inhomogeneity $(\sigma, \delta_0)$ is kept constant while its distance from the source $D$ is increased. We can see that the closer the inhomogeneity, the stronger the strength, which is natural since photons are absorbed and diluted as they travel away from the source. Also, for QSOs, $\ell_{\nu_{1}}$ is huge (cf table \ref{TableMfp}) so that we are always `close' to the source and formulae $B_\text{close}^\text{inside}$ and $B_\text{close}^\text{behind}$ apply. We indeed recover the power-law asymptotic trend with distance of equation \eqref{Bclosebehind}.
In the central panel of figure \ref{Figures_Inhomo}, $D$ and $\delta_0$ are kept constant as $\sigma$ varies. The narrower the density profile, the greater the strength, since gradients are more important for narrow profiles. Note that because the term we are considering in equation \eqref{THE_B} is global, once a non radial gradient is formed, it generates magnetic fields along the entire line of sight. This is why the narrow profiles generate fields of stronger strengths on larger distances. Their angular extent about the symmetry axis is smaller though.
The right panel of \ref{Figures_Inhomo} has to be compared with its central panel: They correspond to the same study but with a MH profile instead of a Gaussian overdensity profile. We can see that the strengths are larger. This comes from the fact that photons in underdensities are less absorbed so that the flux of ionizing photons is more important within the overdensity than in the simply Gaussian overdensity case. This is all the more true than the width of the underdensity is important, which is why contrary to the gaussian case, the wider the profile, the larger the strength for an MH profile.
Also, since the field has opposite directions in underdense and overdense regions, the contributions from the underdensity and the overdensity composing the MH profile cancel out. This is why for such a profile, at some distance, the field changes sign. In the right panel of figure \ref{Figures_Inhomo} we can see that this distance depends on the width of the MH.

In both figures \ref{Figures_Inhomo} and \ref{Figures_Spectra}, when the inhomogeneity is very close to the Str{\"o}mgren sphere, the strength reaches high values at the outer edge of the ionized region. This is because in our model we impose spherically symmetric ionized regions which intercepts the Gaussian profile of the inhomogeneity thus inducing non radial gradients and a magnetic field. In more realistic situations, such border effects will indeed take place as the ionization front will be distorted at the contact of an inhomogeneity. A more precise study of such effects is left for future work.

Finally, expression \eqref{Fglobalgeom} dictates the topology of the magnetic field. Field lines given by $\B \times \vec{dl} = \vec{0}$ in spherical coordinates with $\vec{dl} = dr \ \hat{r} + r d\theta \ \hat{\theta} + r \sin \theta d\phi \ \hat{\phi}$ satisfy
\begin{equation}
dr = 0 \text{ and } \frac{d\theta}{d\varphi} = - \frac{\p_\varphi \int n_{\mathrm{H\textsc{i}}} dr}{\p_\theta \int n_{\mathrm{H\textsc{i}}} dr}.
\label{field_lines}
\end{equation}
Since $dr = 0$, field lines remain on spheres centred on the source. In the case of a $\varphi$-independent configuration, as in all the examples considered here, $d\theta = 0$ so the lines are circles around the axis of symmetry of the system. This is why in all the density plots of this paper the fields point perpendicularly to the plane of the plot. In a more general case the lines remain loops at given radii since $dr = 0$ but with a more complicated shape given by \eqref{field_lines}. Though, once formed, these fields will be processed by the velocity field of the IGM, their peculiar initial spatial configuration is interesting as it may help discriminate magnetic fields generated through this process from those generated by different mechanisms.

\subsubsection{Varying the properties of the source}

In figure \ref{Figures_Spectra} we choose a certain MH inhomogeneity, at $z=10$, and vary the spectral index $\alpha$ and cut-off frequency $\nu_1$ of the source, here a primordial galaxy for illustration.
\begin{figure*}
\includegraphics[scale=0.3]{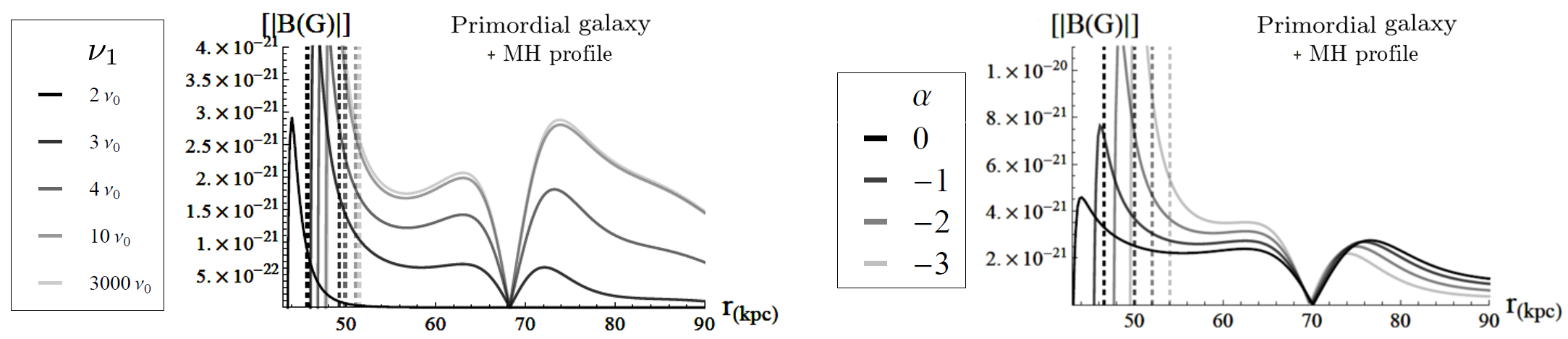}
\caption{Magnetic field generated with various source properties, with a given Mexican hat inhomogeneity at a given position, at $z=10$. Left: For a fixed power-law spectrum, we vary the cut-off frequency $\nu_1$. Right: For a fixed total luminosity, we vary the spectral index $\alpha$. Vertical dashed lines show the distances at which $t_i = t_s$.}
\label{Figures_Spectra}
\end{figure*}

On the left of the figure, we make the cut-off frequency $\nu_1$ vary. We observe that above a certain value, the strength of the generated magnetic field has reached a plateau. For instance in the case shown, the field strength generated with $\nu_1 = 10 \nu_0$ is almost identical to that generated with $\nu_1 = 3000 \nu_0$. This is because $\ell_{10 \nu_0}$ is larger than $50$ kpc at $z=10$, so photons of frequency above $10 \nu_0$ do not interact significantly in the neighbourhood of the inhomogeneity. Also, far behind it they are too diluted to generate a significant magnetic field. This result underlines the following compromise: This mechanism requires photons of high enough energy to photoionize deep in the IGM to generate magnetic fields on large distances, but not too high either because of dilution (cf. equations \eqref{Bclosebehind} and \eqref{Bfarbehind}).

On the right of figure \ref{Figures_Spectra}, we make the spectral index $\alpha$ vary. Since the total luminosity is fixed, harder spectra correspond to sources with fewer low energy photons, which is why harder spectra generate weaker fields near the source, but stronger fields far from the source. Also, naturally, the harder the spectrum the further in the IGM the field is generated. However, we can see in this example that the strengths do not depend too strongly on the spectral index, only up to factors of two. This can be seen in the analytical expressions of the previous section like equation \eqref{InteractionTerm}, where $(\alpha - 5)/3$ and $(\alpha - 6)/3$, for the relevant values of $\alpha$, do not vary much with $\alpha$ and so $B$ does not either.

Finally, we consider various redshifts. Depending on the epoch, the type of ionizing sources and the properties of the IGM are different. For illustration, we decompose the EoR in three stages: PopIII star clusters for $z \in [30,20]$, primordial galaxies for $z \in [20,10]$ and quasars for $z \in [10,6]$. Figure \ref{LapinCretin} shows examples of field configurations obtained at $z=20$ and $z=10$, and table \ref{TableAmplitudes} is a summary of the typical strengths obtained after varying the different parameters of the model.
\begin{figure*}
\vspace{1.5cm}
\includegraphics[scale=0.6]{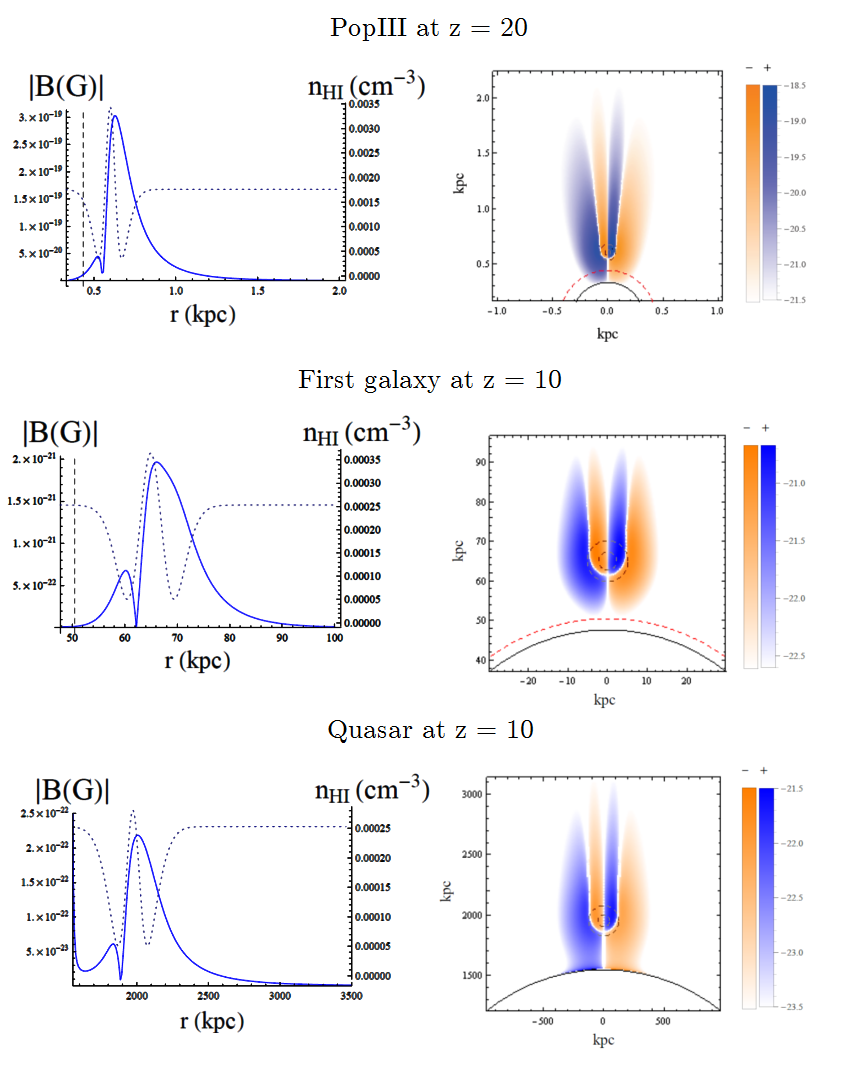}
\caption{Examples of the field generated with Mexican hat inhomogeneities for each type of sources considered here. Left column: Left $y$-axis and blue curves show the absolute value of the magnetic field strength. Vertical dashed lines indicate the distances at which $t_i = t_s$. Right $y$-axis and dotted lines show the inhomogeneity profiles. Right column: Black line is the Str{\"o}mgren sphere. Inner and outer dashed black circles correspond to $\sigma_{+}$ and $\sigma_{-}$ of the inhomogeneity, respectively. Blue areas correspond to magnetic field pointing towards the reader, orange in the opposite direction. Strengths are in Gauss in logarithmic scale.}
\label{LapinCretin}
\vspace{1.5cm}
\end{figure*}
They allow us to identify the following trends. In the beginning of EoR, Population III clusters generate relatively high strengths ($\sim 10^{-19}$ G) but on relatively small scales (hundreds of parsecs), while on the contrary, at the end of EoR, quasars generate low strengths ($\sim 10^{-22}$ G) but on important scales (hundreds of kpc). First galaxies are somewhat in-between and generate $\sim 10^{-21}$ G on tens of kpc in the middle of the EoR.
This may be interpreted as follows. Population III clusters are not very energetic so they form small Str{\"o}mgren spheres and thus photons from the source are not too diluted as they reach the IGM. However, since they appear early in the EoR, $\bar{n}$ is large and thus the mean free path of photons is small. Therefore many photons photoionize and generate strong magnetic fields near the Str{\"o}mgren sphere. On the contrary, quasars are very energetic and generate huge Str{\"o}mgren spheres, so that photons from the source are very diluted as they reach the IGM. In addition since quasars appear late in EoR, the IGM is not very dense and the mean free path of energetic photons is huge. Consequently the ionizing flux is relatively low and the generated magnetic field is weak, but extends on important scales.

\bgroup %change padding in table cells
\def\arraystretch{1} %change padding in table cells

\begin{table}
 \begin{minipage}{86mm}
\footnotesize{
\begin{center}
$\begin{array}{|c|c|c|c|c|}
\hline
\text{Source} & \text{Redshift} & \text{Log } |B| & \text{Distance \ from \ the} \\
 &  & \text{(Gauss)} & \text{ionization \ front \ (kpc)} \\
\hline
\text{Pop III} & 30 & -19 & 0.3 \\
 &  & -21 & 1 \\
 & 20 & -19 & 0.5 \\
 & & -21 & 1 \\
\hline
\text{Primordial} & 20 & -20 & 10 \\
\text{galaxy} &  & -22 & 15 \\
 & 10 & -21 & 30 \\
 & & -22 & 100 \\
\hline
\text{Quasar} & 10 & -21 & 300 \\
 &  & -22 & 1000 \\
 & 6 & -22 & 500 \\
 & & -23 & 1500 \\
\hline
\end{array}$
\end{center}
}

\caption{Third column shows the typical strengths generated at the distances away from the Str{\"o}mgren sphere shown in the fourth column. These strengths are typical values obtained by varying the different properties of the inhomogeneities.}
\label{TableAmplitudes}
\end{minipage}
\end{table}

\section{Discussion}
\label{Discussion}

The cases detailed above are idealized. In reality, as we have mentioned, Str{\"o}mgren spheres are deformed, there is a full distribution of asymmetric inhomogeneities outside the H{\sc ii} regions, and ionizing sources may not be isolated. Taking all this into account would allow us in principle to compute the full magnetic power spectrum generated by photoionization. Furthermore, the IGM is dynamical and the weak turbulence associated with the mildly non linear regime of structure formation on large scales will process the initial power spectrum of the magnetic field. A study of all these effects is beyond the scope of this paper and should be best addressed with dedicated numerical simulations.

As we have seen, an interesting feature of this mechanism is that it generates magnetic fields outside Str{\"o}mgren spheres on distances of the order of a few $\ell_{\nu_1}$, or even larger in underdense regions.
By comparing these distances to the half typical separation between Str{\"o}mgren spheres, we may get an idea of the fraction of the Universe that may be magnetized by this mechanism.
Let us first do so for Population III star clusters at $z = 20$. Considering they formed by molecular cooling in gas overdensities of mass $10^6 M_\odot$ \citep[c.f.][]{Barkana2000}, their half physical mean separation is roughly $10$ kpc, estimated from the abundance of their $\sim 3 \sigma$ parent haloes \citep{Mo2002a}. In addition, the radius of their Str{\"o}mgren spheres is of order a fraction of kpc, and $\ell_{4 \nu_0}$ is about one kpc large. Thus, it seems that these sources magnetize essentially on the outskirts of their Str{\"o}mgren spheres, leaving an important fraction of the IGM unmagnetized.
Consider now faint dwarf galaxies at $z = 15$, and quasars of $10^{12} L_\odot$ at $z = 10$. Faint dwarf galaxies are the major candidate sources responsible for reionization \citep[e.g.][]{Bouwens2012, Wise2014}. Assuming they are hosted in $10^8 M_\odot$ haloes, their half physical mean separation is a couple of tens of kpc. Their Str{\"o}mgren radii are of the order of a few tens of kpc, and $\ell_{4 \nu_0}$ is a couple of kpc large. Luminous quasars, on the other hand, are extremely rare at $z = 10$, which seems confirmed by the recent Planck results implying that the redshift of reionization, $z_\mathrm{re} = 8.8$, is lower than previously thought \citep{Planck2015}. For illustration, considering they are hosted by $5 \sigma$ haloes, their half physical mean separation is of a couple of Mpc. Their Str{\"o}mgren spheres have radii of the order of the Mpc, and these sources have photons of large enough mean free paths to magnetize in between these spheres.
Therefore, these orders of magnitude are more favourable in the case of primordial galaxies and quasars than in the case of Population III star clusters, and suggest that these sources may have participated through this mechanism to the weak magnetization of an important fraction, if not the whole, of the IGM. Note that even the case of Population III star clusters is interesting, as these sources thus premagnetize the environment in which the next generation of stars and galaxies forms.

In this study we have focused on the photoionization term in the induction equation. An exact comparison of this term with the Biermann term is beyond the scope of this paper. We
note that \citet{Doi2011} examined their relative importance in numerical simulations of the neighbourhood of an ionizing super-massive star. In their study, they focus on the situation across an ionization front, in which a self-shielded, neutral, $\delta \simeq 10^2 - 10^3$ over-density defines very sharp and strong gradients in the temperature and electronic density fields. Such a situation could indeed occur within Str\"omgren spheres of the very first luminous sources. Under those conditions, they concluded that the Biermann battery dominates by one order of magnitude. In our case, we considered mild, neutral over-densities way outside the Str\"omgren regions of stronger, long-lived ionizing sources. In such contexts, the Biermann battery may not be effective, be it for purely geometrical reasons. To see that, consider again equation \eqref{induction2}, in which we have separated the global photoionizing contribution from the local contribution. It is the global contribution that creates magnetic fields in regions where the Biermann battery does not operate. Comparison of the two mechanisms is  relevant only in regions where they coexist.
For that purpose, it is convenient to note that equation \eqref{induction2} may be rewritten as
\begin{equation}
\p_t \B = \frac{c}{e} \frac{\vec{\nabla} n_e}{n_e^2} \times \left[ - \vec{\nabla} p_e + \dot{\vec{p}}_\text{pe} \right] - \frac{c}{e n_e} \x \dot{\vec{p}}_\text{pe}.
\label{induction1}
\end{equation}
On the right hand side, the last part contains the full global term examined in this paper (cf. equation \ref{Bglobal}), and a contribution to the local term. The remaining contribution to the local term is contained in the $\dot{\vec{p}}_\text{pe}$ present in the square brackets. In regions where the Biermann battery is efficient, i.e. where pressure and density gradients are not aligned, it is enough to compare  $\dot{\vec{p}}_\text{pe}$ to the pressure gradient. More precisely, since $\dot{\vec{p}}_\text{pe}$ is always radial, their comparison makes sense only in cases where $- \vec{\nabla} p_e$ is radial too.
Then, considering a perfect gas equation of state $p_e = n_e k_b T_e$,  their ratio yields typically
\begin{equation}
\left|\frac{\dot{\vec{p}}_\text{pe}}{n_e k_b \vec{\nabla} T_e}\right| \sim 13
\left(\frac{T_e}{10\, \text{K}}\right)^{-1} \left(\frac{L_g}{10 \,\text{kpc}}\right)
\end{equation}
where $L_g$ is the typical scale of temperature gradients. For illustration, we have taken $\dot{p}_\text{pe} = 3 \times 10^{-44} \ \text{erg} \ \text{cm}^{-4}$, which is obtained at a rather remote distance of $2 r_s \sim 100$ kpc from a primordial galaxy at $z = 10$. Hence, in this case, the photoionization term dominates, even at a distance of $2 r_s$ at which photons are very diluted. Also note that, because the Biermann battery contribution is independent of the properties of the source, the ratio above scales linearly with the luminosity of the ionizing source.

The magnetogenesis mechanism examined here, based on the simple physics of momentum transfer from ionizing photons to photoelectrons, is therefore likely to have generated  seed magnetic fields on cosmological scales during Reionization. Although weak and very remote, the strengths of the seeds produced, together with their specific spatial configuration, could  actually be revealed directly through the recently proposed probe of magnetic fields in the EoR detailed by \citet{Venumadhav2014}, although large coherence lengths of the magnetic fields might be mandatory.

\section*{Acknowledgments}

We would like to thank Nabila Aghanim and Katia Ferri\`ere for stimulating discussions. J.-B. D. acknowledges financial support by the P2IO LabEx (ANR-10-LABX-0038) in the framework `Investissements d'Avenir'  (ANR-11-IDEX-0003-01) managed by the French National Research Agency  (ANR).

\bibliographystyle{mn2e}
\bibliography{IMCD}

\appendix

\section{Momentum transfer term}
\label{App:MomentumTerm}

A fraction $f_\text{mt}$ of the momentum of the incident photon is transferred to the freed electron during photoionization:
\begin{equation}
m_e \vv = f_\text{mt} (\nu) \frac{h \nu}{c} \hat{r},
\label{momentumTransfer}
\end{equation}
where $f_\text{mt}$ is frequency dependent \citep{1930AnP...396..409S}:
\begin{equation}
f_\text{mt} (\nu) = \frac{8}{5} \frac{\nu - \nu_0}{\nu}.
\end{equation}
Note that this fraction may be larger than one, in which case the ions recoil.\\

\noindent Now, by definition $\left.\p_t f_e \right|_s \ddv \ddr dt$ is equal to the number of photoelectrons of speed $v$ in direction $\hat{v}$ generated at a position $\vr$ at a time $t$. Since we consider Hydrogen only, each photoionization produces only one electron. This number is thus equal to the number of photoionizations due to photons of frequency $\nu$ in direction $\hat{k} = \hat{v}$ where $\nu$ satisfies \eqref{momentumTransfer}. Finally, since the photoionization rate density is the product of the number density of incident photons, the velocity of incident photons, the number density of target Hydrogen atoms and the cross section, we get

\begin{equation}
\left.\p_t f_e \right|_s \ddv \ddr dt = \left[ n^\mathrm{inc}_\gamma d\nu d\Omega\right] c \sigma_\nu n_{\mathrm{H\textsc{i}}} \ddr dt
\end{equation}
where the number density of incident photons of frequency $\nu$ with direction $\hat{k}$ at $\vr$ at time $t$ is
\begin{equation}
n^{inc}_\gamma(t,\vr,\hat{k},\nu) = \frac{I_\nu/c}{h \nu}
\end{equation}
by definition of the monochromatic specific intensity. Therefore we model the source term microscopically in equation \eqref{Boltzmann} by
\begin{equation}
\left.\p_t f_e \right|_s \ddv = \frac{I_\nu \sigma_\nu}{h \nu} n_{\mathrm{H\textsc{i}}} d\nu d\Omega,
\label{sourceterm_micro}
\end{equation}
so that macroscopically we get from expression \eqref{sourceterm_macro} from equation \eqref{momentumTransfer_implicit}.

\section{Notations}
\label{App:FluidReduction}
In the multifluid description, macroscopic physical quantities are defined as
\begin{equation}
\begin{array}{l}
n_\alpha = \int f_\alpha \ddv\\
\VV_\alpha = \frac{1}{n_\alpha} \int \vv f_\alpha \ddv\\
P_\alpha = m_\alpha \int \left(\VV_\alpha - \vv\right)\left(\VV_\alpha - \vv\right) f_\alpha \ddv\\
\VJ_\alpha = q_\alpha n_\alpha \VV_\alpha
\end{array}
\end{equation}
respectively the number density, the velocity, the pressure tensor and the current density of species $\alpha$. This description is then reduced to a single fluid with
\begin{equation}
\begin{array}{l}
\rho_m = \Sigma_\alpha n_\alpha m_\alpha\\
\rho_q = \Sigma_\alpha n_\alpha q_\alpha\\
\VV = \frac{\Sigma_\alpha n_\alpha m_\alpha \VV_\alpha}{\Sigma_\alpha n_\alpha m_\alpha}\\
\VJ = \Sigma_\alpha \VJ_\alpha\\
P = \Sigma_\alpha P_\alpha^{CM}\\
P_\alpha^{CM} = m_\alpha \int \left(\VV - \vv\right)\left(\VV - \vv\right)f_\alpha \ddv
\end{array}
\end{equation}
respectively the mass density, the charge density, the centre-of-mass velocity, the current density and the total centre-of-mass pressure tensor in the one-fluid. Then, taking the first moment of \eqref{Boltzmann} weighted by $q_\alpha$, and summing over all species, yields the generalized Ohm's law \eqref{OhmFull}, with the additional notations:
\begin{equation}
\begin{array}{l}
\vec{P} \equiv \Sigma_\alpha \frac{q_\alpha}{m_\alpha} \dv P_\alpha^{CM}\\
\vec{C} \equiv \Sigma_\alpha q_\alpha \int \vv \left.\p_t f_\alpha \right|_c \ddv.
\end{array}
\end{equation}

\bsp

\label{lastpage}

\end{document}